\documentclass[journal]{IEEEtran}
\usepackage{amsmath,amssymb,amsfonts,mathrsfs,mathtools,bm, bbm, dsfont,mathrsfs,amsthm,blkarray}
\usepackage[dvipdfm]{graphicx}
\usepackage[dvipsnames]{xcolor}

\usepackage{epsfig,epsf,psfrag,latexsym}

\usepackage{booktabs}
\usepackage{multirow,multicol}

\usepackage[breaklinks,colorlinks, linkcolor=MidnightBlue, anchorcolor=MidnightBlue, citecolor=MidnightBlue, urlcolor=MidnightBlue]{hyperref}
\usepackage[hyphenbreaks]{breakurl}
\usepackage{cite}
\usepackage{xurl}

\newcommand*{\QED}{\hfill\ensuremath{\square}}

 \def\old#1{}    

\def\nn{\nonumber}
\def\beq{\begin{equation}}
\def\eeq{\end{equation}}
\def\bea{\begin{eqnarray}}
\def\eea{\end{eqnarray}}
\def\ba{\begin{array}}
\def\ea{\end{array}}

\def\bitem{\begin{itemize}}
\def\eitem{\end{itemize}}
\def\ben{\begin{enumerate}}
\def\een{\end{enumerate}}

\def\eg{{\it e.g., \/}}

\def\ie{{\it i.e.,\ \/}}

\def\ebf{{\bm e}}

\def\gbf{{\bm g}}

\def\mbf{{\bm m}}

\def\qbf{{\bm q}}
\def\rbf{{\bm r}}
\def\sbf{{\bm s}}

\def\ubf{{\bm u}}

\def\xbf{{\bm x}}

\def\rbf{{\bm r}}
\def\xbf{{\bm x}}

\def\Bbf{{\bm B}}

\def\Ebf{{\bm E}}

\def\Rbf{{\bm R}}
\def\Sbf{{\bm S}}

\def\Ec{{\cal E}}

\newcommand{\beqa}{\begin{eqnarray}}
\newcommand{\eeqa}{\end{eqnarray}}
\newcommand{\beqan}{\begin{eqnarray*}}
\newcommand{\eeqan}{\end{eqnarray*}}

\newcounter{l1}
\newcounter{l2}
\newcounter{l3}
\newcommand{\bdotlist}{\begin{list}{$\bullet$}{}}
\newcommand{\bboxlist}{\begin{list}{$\Box$}{}}
\newcommand{\bbboxlist}{\begin{list}{\raisebox{.005in}{{\tiny
$\blacksquare$ \ \ }}}{}}
\newcommand{\bdashlist}{\begin{list}{$-$}{} }
\newcommand{\blist}{\begin{list}{}{} }
\newcommand{\barablist}{\begin{list}{\arabic{l1}}{\usecounter{l1}}}
\newcommand{\balphlist}{\begin{list}{(\alph{l2})}{\usecounter{l2}}}
\newcommand{\bAlphlist}{\begin{list}{\Alph{l2}.}{\usecounter{l2}}}
\newcommand{\bdiamlist}{\begin{list}{$\diamond$}{}}
\newcommand{\bromalist}{\begin{list}{(\roman{l3})}{\usecounter{l3}}}

\newtheorem{theorem}{Theorem}
\newtheorem{lemma}{Lemma}
\newtheorem{proposition}{Proposition}

\newtheorem{assumption}{Assumption}

\title{Convexifying Regulation Market Clearing of State-of-Charge Dependent Bid}

\author{Siying Li~\IEEEmembership{Student Member,~IEEE,}
\quad
Cong Chen ~\IEEEmembership{Student Member,~IEEE,}
\quad Lang~Tong~\IEEEmembership{Fellow,~IEEE}
\thanks{\scriptsize
Siying Li, Cong Chen,  and Lang Tong (\{sl2843, cc2662, lt35\}@cornell.edu) are with the School of Electrical and Computer Engineering, Cornell University, Ithaca NY, USA. (Corresponding author: Cong Chen)}
\thanks{\scriptsize This work was supported in part by the National Science Foundation under Award 2218110 and the Power Systems and Engineering Research Center (PSERC) Research Project M-46.}
}

\begin{document}
\maketitle

\begin{abstract}

We consider the problem of merchant storage
participating in the regulation market with state-of-charge (SoC) dependent bids. Because storage can simultaneously provide regulation up and regulation down capacities, the market-clearing engine faces the computation challenge of evaluating storage costs under different regulation scenarios. One approach is to employ a bilevel optimization that minimizes the worst-case storage cost among all potential regulation events. However, subproblems of such a bilevel optimization are nonconvex, resulting in prohibitive computation challenges for the real-time clearing of the regulation market. We show that the complex nonconvex market clearing problem can be convexified by a simple restriction on the SoC-dependent bid, rendering the intractable market clearing computation to standard linear programs. Numerical simulations demonstrate that SoC-dependent bids satisfying the convexification conditions increase the profits of merchant storage owners by 12.32-77.38\% compared with SoC-independent bids.  
\end{abstract}


\section{Introduction}\label{sec:Intro}
With the increasing penetration of renewables in bulk transmission systems, participation of merchant battery energy storage systems (BESS) at scale is essential in mitigating volatilities of renewable generation, reducing clean energy curtailment, and minimizing the dependency on thermal generation for ramping support.  To this end, an efficient and computationally scalable market-clearing engine is crucial.

Unlike traditional generation and demand resources, BESS's operating costs consist of battery degradation and opportunity costs, both tied to BESS's State-of-Charge (SOC) trajectories. Therefore, it is natural that merchant BESS participants and the market operator are interested in including SoC-dependent bids and offers in the energy and regulation markets \cite{CAISO_SOCdependent:22, REV_SOCdependent:22}.

A SoC-dependent bid specifies the BESS charging and discharging bid-in costs as a function of BESS SoC, typically in the standard parametric piecewise linear form.  For a market clearing optimization involving SoC-dependent bid-in costs, the standard economic dispatch problem becomes non-convex. Although mixed-integer linear programs (MILP) have been proposed to clear SoC-dependent bids \cite{ZhengXu22energy, ZhengXu22socAribitrage}, the computation cost of MILP is nontrivial for real-time energy market and is a major barrier preventing the adoption of SoC-dependent bids in real-time market operations.

A promising recent approach \cite{ChenTong22arXivSoC} to overcome computational challenges of clearing SoC-dependent bids in energy market is imposing the so-called {\em equal decremental-cost ratio (EDCR)} condition on the bidding format. With the EDCR bidding, the MILP energy market clearing becomes a standard linear program\footnote{In \cite{ChenTong22arXivSoC}, the energy market clearing under EDCR condition is transformed into a convex optimization with piecewise linear objective and linear constraints, which can be easily rewritten into a linear program.} that can be solved easily by an existing market clearing engine. The impacts of EDCR restriction on SoC-dependent bid is unknown, however, especially on whether the EDCR SoC-dependent bid undermines the profitability of merchant BESS.

This work extends the idea of convexifying the market-clearing of SoC-dependent bids in the frequency regulation market where BESS participants are currently most profitable. At the outset, the market clearing problems are significantly different for energy and regulation markets. First, the energy market clearing {\em prohibits simultaneous charging and discharging} of a battery, whereas the regulation market clearing {\em allows simultaneously clearing the up and down regulation capacities}\cite{CAISO_StakeholderComments:22}. Second, the minute-level clearing of real-time regulation market\footnote{The real-time regulation capacity clearing operates at 15-minute intervals in CAISO \cite{CAISO_RegulationMarket:22}.} requires evaluating costs associated with the up and down-regulation capacities, where complications arise when such costs are functions of the BESS SoC trajectory set by the fast timescale regulation signals at, \eg the four-second timescale. These major differences render the convexification techniques used in the energy market inadequate for the regulation market.

\vspace{-0.1in}
\subsection{Summary of contribution}
This paper proposes a novel convex market-clearing solution for the frequency regulation market involving SoC-dependent bids. The main contribution is three-fold.

First, we present a rigorous formulation to clear the regulation market with SoC-dependent bid. We allow BESS to simultaneously provide regulation up and down capacities and consider the effects of all possible SoC trajectories set by future regulation signals. Such a formulation leads naturally to a bilevel optimization that determines the worst trajectories with the highest up and down-regulation BESS costs.

Second, we show in Theorem~\ref{thm:ROCVX} that, although the bilevel optimization involved in the market clearing is non-convex in general, the EDCR bids proposed in \cite{ChenTong22arXivSoC} convexifies the bilevel non-convex problem to a standard linear program, resulting in the use of standard market clearing engine for SoC-dependent bids in frequency regulation markets. This theoretical result is nontrivial, surprising, and significant.

Finally, we examine the benefits of the SoC-dependent bid through numerical simulations, demonstrating that, when compared with {\em SoC-independent bids}, {\em SoC-dependent bids} increase storage participants' profits and decrease overall system operation costs. In particular, with SoC-dependent bids, storage profits increased by approximately 12.32-77.38\%, and system cost decreased by 1.38-4.17\% in our simulation.

\vspace{-0.1in}
\subsection{Paper organization and notations}
The paper is organized as follows. In Sec.~\ref{sec:A_Access}, we first introduce the SoC-dependent bid in the regulation market, and then we establish a bilevel optimization to clear the SoC-dependent bid in the regulation market. We show that the nonconvex subproblems, computing the SoC-dependent cost of storage, makes the bilevel optimization intractable. In Sec.~\ref{sec:AccessRight}, we present the convexification procedure for this nontrivial bilevel optimization to clear SoC-dependent bids in the regulation market. Related simulations are in Sec.~\ref{sec:SOAccessRight}. 

For the notations, we use $x$ for scalar, $\xbf$ for vector, and define $[x]:=\{1,...,x\}$. The indicator function $\mathbbm{1}$ means that  $\mathbbm{1}_{\cal X}=1$ if   ${\cal X}$ is true, and $\mathbbm{1}_{\cal X}=0$ otherwise. All superscripts in this paper are used to distinguish notations rather than representing the power of a number.

\section{  Nonconvex SOC-dependent regulation market  }\label{sec:A_Access}
 So far, various regulation market operation methods have been proposed. Pay-for-performance regulation markets are among the most common, operating by clearing both regulation capacity and regulation mileage\cite{Chen:15}. Some markets include both regulation up and regulation down services, while others not \cite{Argonne:16}. For the regulation market allowing unidirectional regulation capacity, the approach in our earlier work \cite{ChenTong22arXivSoC} can be easily extended from the energy market to the regulation market with SoC dependency. However, for storage providing bidirectional regulation capacities, the computation of SoC-dependent cost becomes nontrivial.

\vspace{-0.1in}
\subsection{BESS bids in the regulation market} \label{2-A}
The regulation market and energy market are jointly cleared with 15-minute as the time granularity \cite{CAISO_RegulationMarket:22}. We consider BESS only participating in the regulation market and providing regulation up and regulation down capacities denoted by $r^u_{t}$ and $r^d_{t}$, respectively, at time $t$. Following the formula in CAISO\cite{CAISO_SOCImplementation:23}\footnote{The SoC evolving with $e_{t+1}=e_{t} + \gamma^d_{t}r^d_{t}\eta-\gamma^u_{t}r^u_{t}$ in \cite{CAISO_SOCImplementation:23}. $\gamma^d_{t}$ and $\gamma^u_{t}$ are the expected values of the multipliers that indicate the fraction of regulation up and down capacity that is actually utilized.  We ignore these pre-assigned coefficients in our model for simplicity.}, the SoC $e_t$ evolves based on  
\beq
\begin{array}{lcl}\label{eq:SoCevolveReg}
e_{t+1}=e_{t} + r^d_{t}\eta-r^u_{t}, 
\end{array}
\eeq
where $\eta$ is the round trip efficiency of the storage, and the units for regulation capacities and SoC are all MWh. The storage provides regulation down capacities with its charging energy, and discharges for regulation up. Note that in the regulation market, storage can provide both regulation up and regulation down capacity within the same 15-min interval \cite{CAISO_StakeholderComments:22}. To ensure that the storage behaves in a physically realistic manner during intervals with bidirectional regulation capacities cleared, the operation limits for SoC are given by
\beq\label{eq:SoClimit}
\begin{array}{cc}
     &e_{t}+ r^d_{t}\eta \leq \overline{E},\\
     & \underline{E} \le e_{t}- r^u_{t},\\
\end{array}
\eeq
where $\underline{E}$ and $\overline{E}$ are  minimum and maximum SoC limits.

Regulation capacity refers to the reservation of a specific capacity to facilitate real-time frequency regulation service, it does not present the realization \cite{Chen:15}. The actual manifestation of storage operation in the regulation market is correlated with the regulation mileage, which follows the Automatic Generation Control (AGC) signal delivered every 4-s.  During the 15-min interval indexed by $t$, we denote $m^u_{j,t}, m^d_{j,t}, \hat{e}_{j,t}$ as the regulation up mileage, regulation down mileage and SoC at the $j$-th 4-s interval.  The storage SoC $\hat{e}_{j,t}$ evolves with 
\beq\label{eq:4sSOC}
\begin{array}{cc}
     &\hat{e}_{j,t} := e_{t}+ (\sum_{i=1}^{j-1}m^d_{i,t}\eta-\sum_{i=1}^{j-1} m^u_{i,t})\delta,\\
     & m^u_{j,t} m^d_{j,t}=0,\\
\end{array}
\eeq
where we use $\delta$ representing the time length of  4-s interval. The unit for regulation up/down mileage is MW. AGC won't require storage to simultaneously regulate up and down\cite{ERCOTReg:23}.

 \begin{figure}[!htb]
   \centering
   \vspace{-0.1in}
   \includegraphics[scale=1.6]{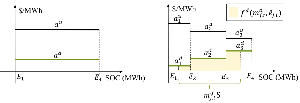}
\caption{\scriptsize BESS bids in the regulation market. Left: SoC-independent bid. Right: The SoC-dependent bid format when $K=3$. Cost of the storage initialing at $\hat{e}_{j,t}$ and providing regulation down mileage by $m^d_{j,t} $ during time $\delta=$ 4s. }
\label{fig:SOC_D_Bid}
\end{figure}
 \vspace{-0.1in}
BESS bid in the current regulation market for regulation capacity with SoC-independent bid, as shown on the left of Fig.~\ref{fig:SOC_D_Bid}. The proposed SoC-dependent bid for regulation capacity is shown on the right of Fig.~\ref{fig:SOC_D_Bid}, which is an extension of the idea in \cite{CAISO_SOCdependent:22} for the energy market.  Without loss of generality, we partition the SoC axis into $K$ consecutive segments, within each segment $\Ec_k:=[E_k,E_{k+1}]$, a pair of bid-in {\em marginal cost} parameters  $(a_k^u,a_k^d)$ for regulation up and regulation down capacites is defined. 

SoC-dependent bids enhance the longevity of the battery and the storage's ability to capture profit opportunities in the market.  When the SoC is low, it is more costly to discharge while less costly to charge. As a result, typical bid-in regulation up costs $(a^u_{k})$ monotonically decrease, while regulation down costs $(a^d_{k})$ monotonically increase. Therefore, SoC-dependent bids satisfy the following conditions.
 
\begin{assumption}[] \label{assume:single} The SoC-dependent cost parameters $(a_k^u,a_k^d)$ satisfy  $0 \leq a_{k+1}^u \leq a_k^u, 0\le a_k^d \le a^d_{k+1}, \forall k \in [K-1]$.
\end{assumption}

\subsection{SoC-dependent bid-in cost in the regulation market}


 SoC-dependent bids induce SoC-dependent scheduling costs for storage. In the regulation market, the SoC-dependent cost for storage is dependent on the regulation mileage. Denote $J$ as the number of 4-s time intervals within interval $t$, $\mbf^u_t:=(m_{1,t}^u,...,m_{j,t}^u,...,m_{J,t}^u)$, and similarly for $\mbf^d_t$. The SoC-dependent cost during a 15-min interval $t$ is the sum of all single stage cost following the 4-s AGC signals, given by
\beq \label{eq:SoCcostMile}
f^b(\mbf^u_t,\mbf^d_t,e_t):=  \sum_{j=1}^J (f^u(m^u_{j,t}, \hat{e}_{j,t} )+f^d(m^d_{j,t}, \hat{e}_{j,t} )),
\eeq
where $f^u$ and  $f^d$ are the SoC-dependent cost for regulation up and regulation down respectively during a 4-s interval. The SoC-dependent scheduling costs involving the ({\it ex ante}) SoC $\hat{e}_{j,t}$ at the beginning of the $j$-th 4-s interval and the ({\it ex post}) SoC $\hat{e}_{j+1,t}$ at the end of the $j$-th interval, which may be in different SoC partitioned segments. We denote $m$ and $n$ as index for the SoC-segment with $\hat{e}_{j,t} \in \Ec_m, \hat{e}_{j+1,t}\in \Ec_{n}$. The regulation down cost $f^d$ is given by 
\beq
f^d(x,y):=\mathbbm{1}_{\{n\geq m\}} a^d_{n}\delta x+\mathbbm{1}_{\{n > m\}}\sum_{k=m}^{n-1}\frac{\Delta a^d_{k}}{\eta}(E_{k+1}-y),\nn
\eeq
where $x$ and $y$ are used as symbols rather than variables with specific physical interpretations, $\Delta a^u_{k}:=a^u_{k}-a^u_{k+1}$ and $\Delta a^d_{k}:=a^d_{k}-a^d_{k+1}$. The yellow area on the right of Fig.~\ref{fig:SOC_D_Bid} represents the SoC-dependent cost when the storage is initialized at SoC $\hat{e}_{j,t}$ and provides regulation down mileage $m_{j,t}^d$ during time $\delta$. In parallel, the SoC-dependent cost when providing regulation up mileage at time $j$ is given by 
\beq
f^u(x,y):=\mathbbm{1}_{\{n\le m\}} a^u_{n} \delta x+\mathbbm{1}_{\{n < m\}}\sum_{k=n+1}^{m} \Delta a^u_{k-1}(E_{k}-y).\nn
\eeq

 


\vspace{-0.2in}
\subsection{Bilevel optimization in the regulation market}

The regulation market is responsible for clearing the regulation capacities based on bid-in parameters from BESS. Although the SoC-dependent cost can be accurately captured by \eqref{eq:SoCcostMile}, the regulation mileages are unknown when clearing SoC-dependent bids in the regulation market ahead of time. We only know that AGC signals are required to stay within the assigned regulation capacities, \ie
\beq\label{eq:AGC}
    \begin{array}{l}
    \sum_{j=1}^Jm^u_{j,t}\delta \le  r^u_t, \\\sum_{j=1}^J m^d_{j,t}\delta \le r^d_t.
\end{array}
\eeq
Without detailed 4-s-based SoC trajectories inside the 15-min interval $t$, precise computation of the SoC-dependent cost is challenging when clearing the regulation capacity ex-ante. To address this issue, we propose a bilevel  optimization to clear the SoC-dependent regulation market. The upper level model \eqref{eq:NONCVX_REG_UP} aims to minimize the system operation cost with a time granularity of 15-min to clear the regulation capacities. The lower level model \eqref{eq:ROmileage} is employed to obtain the worst-case SoC-dependent storage cost with a time granularity of 4-s.

Inside each 15-min interval, we identify the worst-case SoC-dependent cost, given the regulation capacity, with 

{\em Lower level optimization:}
\beq
\label{eq:ROmileage}
    \begin{array}{lrl}
f^*(r^u_t, r^d_t,e_t) = &\underset{ \mbf^u_t,\mbf^d_t,\hat{\ebf}_t\in \Rbf^J_+}{\rm maxmize}& f^b(\mbf^u_t,\mbf^d_t,e_t) \\
&\mbox{subject to}& \forall j \in [J], \eqref{eq:4sSOC}, \eqref{eq:AGC},
\end{array}
\eeq
which is nonconvex because of two reasons: (i) the storage cost  in objective function is computed by \eqref{eq:SoCcostMile}, a nonconvex function; (ii) The constraint \eqref{eq:4sSOC} that prohibits storage from simultaneously exhibiting regulation up and down mileage is bilinear. The limits for regulation mileage are disregarded because we assume storage has fast charging/discharging ability.\footnote{We ignore the SoC capacity limits of $\overline{E}$ and $\underline{E}$ as \eqref{eq:SoClimit} in this model since it is guaranteed to be feasible when assigning the regulation capacities.}


Solving the above optimization yields the optimal decisions $\mbf^{u*},~\mbf^{d*}$, and the optimal value for the worst-case storage SoC-dependent cost $f^*(r^u_t, r^d_t,e_t)$ in time $t$. So, considering $T$ segments of the 15-min interval when clearing regulation capacities, the $T$-interval SoC-dependent storage cost is  
\beq\label{eq:robustcost}
F^*(\rbf^u, \rbf^d;s) =\sum_{t=1}^T f^*(r^u_t, r^d_t,e_t) .
\eeq





With the robust SoC-dependent cost \eqref{eq:robustcost} for storage, we establish the energy-regulation co-optimization to clear SoC-dependent bids for the regulation market. The storage index is omitted in previous sections for simplicity. We introduce the storage index $i$ in the following model, which includes multiple storage in the market clearing.

{\em Upper level optimization:}
\begin{subequations}
\label{eq:NONCVX_REG_UP}
\begin{align}
& \underset{\substack{\{\gbf_i^e, \gbf_i^u, \gbf_i^d,\\\pmb{r}_{i}^u,\pmb{r}_{i}^d \in \Rbf^T_+ \}}}{\rm minimize} && \sum_{i=1}^{M} (h_i(\gbf^e_i, \gbf_i^{u}, \gbf_i^{d})+ F^*_{i}(\rbf^u_{i}, \rbf^d_{i};s_i))\label{eq:obj} \\
& \mbox{subject to}&& \forall t\in [T], \forall i\in [M],\nn \\ \label{eq:PFlimit}
&&& \pmb{S} (\pmb{g}^e[t]-\pmb{d}[t]) \le \pmb{q},\\ \label{eq:PF}
&&& {\bf 1}^\intercal(\pmb{g}^e[t]-\pmb{d}[t]) = 0,\\ \label{eq:ReqReg1}
&\beta^u_{t}: && \sum_i^M (g_{it}^{u} + r_{it}^u) \geq \xi_t^u, \\ \label{eq:ReqReg2}
&\beta^d_{t}: && \sum_i^M (g_{it}^{d} + r_{it}^d) \geq \xi_t^d, \\ \label{eq:gen}
&&& g^e_{it}+g^u_{it}\leq \bar{g}_i,~~ \underline{g}_i \le g^e_{it}-g^d_{it}, \\ \label{eq:SoCcons1}
&&& e_{i1}=s_i,  ~~\eqref{eq:SoCevolveReg}, ~~\eqref{eq:SoClimit},\\ \label{eq:SoCcons2}
&&& 0\le r^d_{it}\le \bar{r}^d_i, ~~0\le r^u_{it}\le \bar{r}^u_i,
    \end{align}
\end{subequations}
where  $h_i(\gbf^e_i, \gbf_i^{u}, \gbf_i^{d})$ in \eqref{eq:obj} represents the convex generator cost, $\gbf^e_i$, $\gbf_i^{u}$, and $\gbf_i^{d}$ denote the generation for energy market, the regulation up capacity, and the regulation down capacity of generator $i$. We consider a $M$-bus system, in which each power network bus has one generator. Storage only participates in the regulation market. DC power flow model is adopted in \eqref{eq:PFlimit}-\eqref{eq:PF} with notations  $\Sbf \in \mathbb{R}^{2B\times M}$ for the shift-factor matrix. $B$ is the number of branches in the network  and the branch flow limit is $ \qbf \in \mathbb{R}^{2B}$. Regulation capacity requirements \eqref{eq:ReqReg1}-\eqref{eq:ReqReg2}, and generator capacity limits \eqref{eq:gen} are considered given zonal regulation requirements $\xi_t^u$ and $\xi_t^d$. $\beta^u_{t}, \beta^d_{t}$ are dual variables defined correspondingly.  The storage constraints in \eqref{eq:SoCcons1}-\eqref{eq:SoCcons2} include SoC transition constraints, SoC limits, and regulation up/down capacity limits. The  initial SoC $s_i$ is given.
 
Such a bilevel  optimization is intractable because of the nonconvexity of the lower level subproblem \eqref{eq:ROmileage}, imposing a significant computational burden on the real-time electricity market. To enable an efficient energy-regulation co-optimization in the real time, we propose a convexification condition in the following section. 

\section{Convex SoC-dependent regulation market}\label{sec:AccessRight}

In this section, we introduce the equal decremental-cost ratio (EDCR) condition imposed on the bid-in cost parameters to convexify the SoC-dependent regulation market. Storage index $i$ is omitted in this section for simplicity.

\vspace{-0.1in}
\subsection{EDCR condition in the regulation market}
Although the bilevel optimization in Sec.~\ref{sec:A_Access} is in general nontrivial, we discover a simple and sufficient condition to achieve a convex analytical form for the optimal value of the lower level subproblem \eqref{eq:ROmileage}. From that, we reformulate the bilevel optimization into a single level convex optimization to clear SoC-dependent bids in the regulation market. This is summarized in the following theorem.
 
\begin{theorem}[] \label{thm:ROCVX} If  storage's bid-in parameters satisfy the equal decremental-cost ratio (EDCR) condition,
\begin{equation}\label{eq:EDCR}
\frac{a^d_{k-1}-a^d_{k}}{a^u_k-a^u_{k-1}}=\eta, \forall k, 
\end{equation}
then, under Assumption~\ref{assume:single}, the robust  SoC-dependent cost \eqref{eq:robustcost} can be computed by a convex piecewise linear function
\beq
\begin{array}{lrl}\label{eq:ESROcost}
\tilde{F}(\rbf^u, \rbf^d; s) =\underset{j\in [K]}{\rm max}\{\alpha_j(s) + \sum_{t=1}^T(a^d_{j}r_t^d+a^u_{j}r_t^u)\},
\end{array}
\eeq
where $\alpha_j(s):=\sum_{k=1}^{j-1}\frac{\Delta a^d_k(E_{k+1}-E_1)}{\eta} + \frac{a^d_{j}(s-E_1)}{\eta}+h(s)$ and 
$h(s):=\sum_{i=1}^K\mathbbm{1}_{\{s\in \Ec_i\}}(\frac{a^d_i(E_1 - s)}{\eta} - \sum_{k=1}^{i-1}\frac{\Delta a^d_k (E_{k+1}-E_1)}{\eta})$.
\end{theorem}

The proof can be found in Appendix~\ref{sec:THM1}. So under EDCR condition, we can substitute the storage cost in (\ref{eq:obj}) with \eqref{eq:ESROcost}, and reformulation \eqref{eq:NONCVX_REG_UP} into a standard linear program (See Appendix~\ref{sec:LP} for the linear program reformulation). This allows using a standard market clearing engine for SoC-dependent bids in frequency regulation markets. This theorem is in parallel with our previous paper \cite{ChenTong22arXivSoC}, where the EDCR condition is first proposed to convexify the  energy market clearing problem with SoC-dependent bid. The extension to the regulation market here is surprising and challenging since the regulation market clearing of SoC-dependent bids is a complicated bilevel optimization. 

\vspace{-0.1in}
\subsection{Convex SoC-dependent regulation market } 

With the convex storage cost in Theorem~\ref{thm:ROCVX}, we can substitute the storage cost in (\ref{eq:obj}) with \eqref{eq:ESROcost}, making the market clearing of the SoC-dependent regulation market convex. After solving this convex optimization, we obtain the optimal solution for primal variables,\ie $r^{u*}_{t}$ and $r^{d*}_{t}$, and dual variables, \ie $\beta^{u*}_t$ and $\beta^{d*}_t$, which are defined as the prices for  regulation up and regulation down capacities at interval $t$, respectively.
Based on the dispatch results $r^{u*}_{t}$ and $r^{d*}_{t}$, the storage receives payment for supplying regulation capacity according to
\beq
{\cal P}(\rbf^{u*}, \rbf^{d*}) = \sum_{t=1}^T (  \beta^{u*}_{t}r^{u*}_{t}+ \beta^{d*}_{t}r^{d*}_{t}).
\eeq

The bid-in profit calculated based on the bid-in cost  \eqref{eq:ESROcost} is 
\beq
\tilde{\Pi}(\rbf^{u*}, \rbf^{d*};s) = {\cal P}(\rbf^{u*}, \rbf^{d*})-\tilde{F}(\rbf^{u*}, \rbf^{d*};s).
\eeq

Since storage may submit bid-in cost $\tilde{F}$ different from its true cost $F$, the true profit of storage is given below 
\beq\label{eq:true cost}
\Pi(\rbf^{u*}, \rbf^{d*};s) = {\cal P}(\rbf^{u*}, \rbf^{d*})-F(\rbf^{u*}, \rbf^{d*};s),
\eeq
which may also differ from its bid-in profit $\tilde{\Pi}$.

Intuitively, storage won't offer bidirectional regulation capacities when the clearing prices in the regulation market are lower than the bid-in cost of storage to provide bidirectional regulation capacities. This is summarized in the following.
\begin{proposition}\label{Prop:Simultaneous} Under Assumption~\ref{assume:single}, if the SoC-dependent cost parameters $(a_k^u,a_k^d)$ satisfy the EDCR condition in \eqref{eq:EDCR} and
\beq\label{eq:unidirectional}
a^u_K\eta+a^d_1>\beta^{u*}_t\eta+\beta^{d*}_t,
\eeq 
then the storage has unidirectional regulation capacities cleared over all time intervals, \ie $r^{u*}_t r^{d*}_t = 0, \forall t$. 
\end{proposition}

The proof of this proposition can be found in Appendix~\ref{sec:prop1}. In such scenarios, the cleared regulation capacity is unidirectional, reducing the complexity of computing SoC-dependent costs in the regulation market. 

We also propose a heuristic approach to approximate the worst-case storage cost in \eqref{eq:ROmileage} without any restrictions, such as \eqref{eq:EDCR} and
\eqref{eq:unidirectional}. We reformulate the regulation market clearing \eqref{eq:NONCVX_REG_UP} as a mixed-integer program (MIP) using this heuristic approach. The detailed MIP formulation is shown in the Appendix~\ref{sec:MIPheuristic}. Although still nonconvex, this MIP avoid dealing with the intractable bilevel model. In scenarios where unidirectional regulation capacities are cleared, such a heuristic is accurate.


\section{Simulation}\label{sec:SOAccessRight}
\begin{figure}[!htb]
   \centering
   \vspace{-0.1in}
   \includegraphics[scale=1.2]{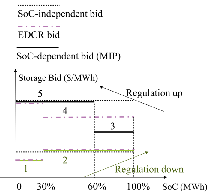} \includegraphics[scale=1.2]{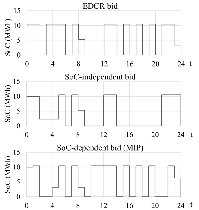} 
\caption{\scriptsize Left: bidding parameters for storage; Right: storage SoC trajectory in the one-shot dispatch under one random scenario.}
\label{fig:ESbid}
\end{figure}

We employed a {\em three-generator case} to demonstrate how storage participates in the regulation market.  It's a single bus system that consists of 2 conventional generators, 1 wind generator with variable output, and 1 ideal energy storage\footnote{The round trip efficiency for ideal storage equals 1.}. The left of Fig.~\ref{fig:ESbid} shows storage bids in three cases. The EDCR bid and SoC-independent bid were generated via approximating the true SOC-dependent bid (solid line) with the method in \cite{ChenTong22arXivSoC}, \cite{ChenLiTong23SoC}.\footnote{In the left of Fig.~\ref{fig:ESbid}, the EDCR bid is closer to the true SoC-dependent bid than the SoC-independent bid.} We simulated the case when storage has unidirectional regulation capacities cleared. The MIP (in Appendix~\ref{sec:MIPheuristic}) was adopted to conduct the regulation market clearing for the true SoC-dependent bid. Convex optimizations are adopted to clear the EDCR bid and SoC-independent bid.


 
We sampled 100 random scenarios with the wind velocity at time $t$ following a Gaussian distribution ${\cal N}(\mu_t, 5)$ for the wind generator. The mean value of wind velocity $\pmb{\mu}$ was given for each interval during the 24-T period\footnote{Here, $T$ represents the 15-min interval.}, which was detailed in the appendix. The wind generator's output had a capacity limit of 20 MW, and the wind generation model was referred from \cite{Giorsetto:83}. The other two generators had regulation up/down capacity limits of 15 MW and 40 MW, respectively. The storage with capacity $\overline{E}=10.5$ MWh had initial SoC $s=10$ MWh.

\begin{figure}[!htb]
   \centering
   \vspace{-0.1in}
\includegraphics[scale=0.23]{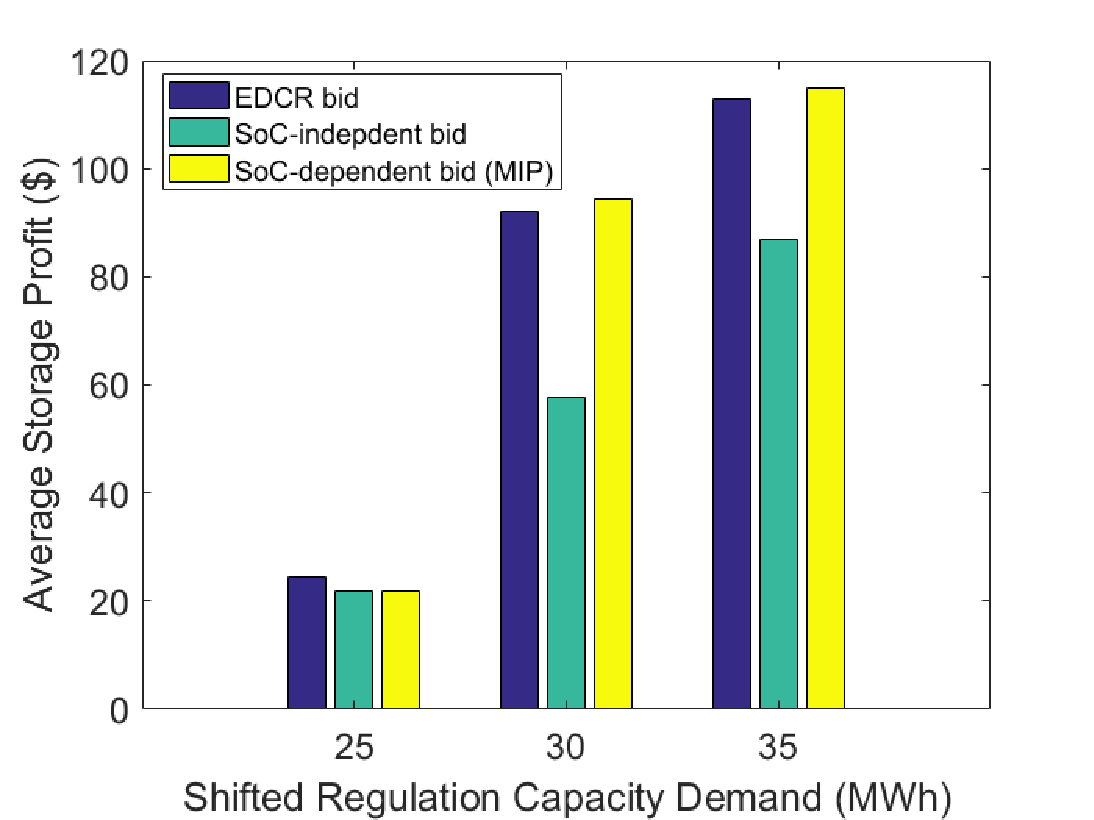}   \includegraphics[scale=0.23]{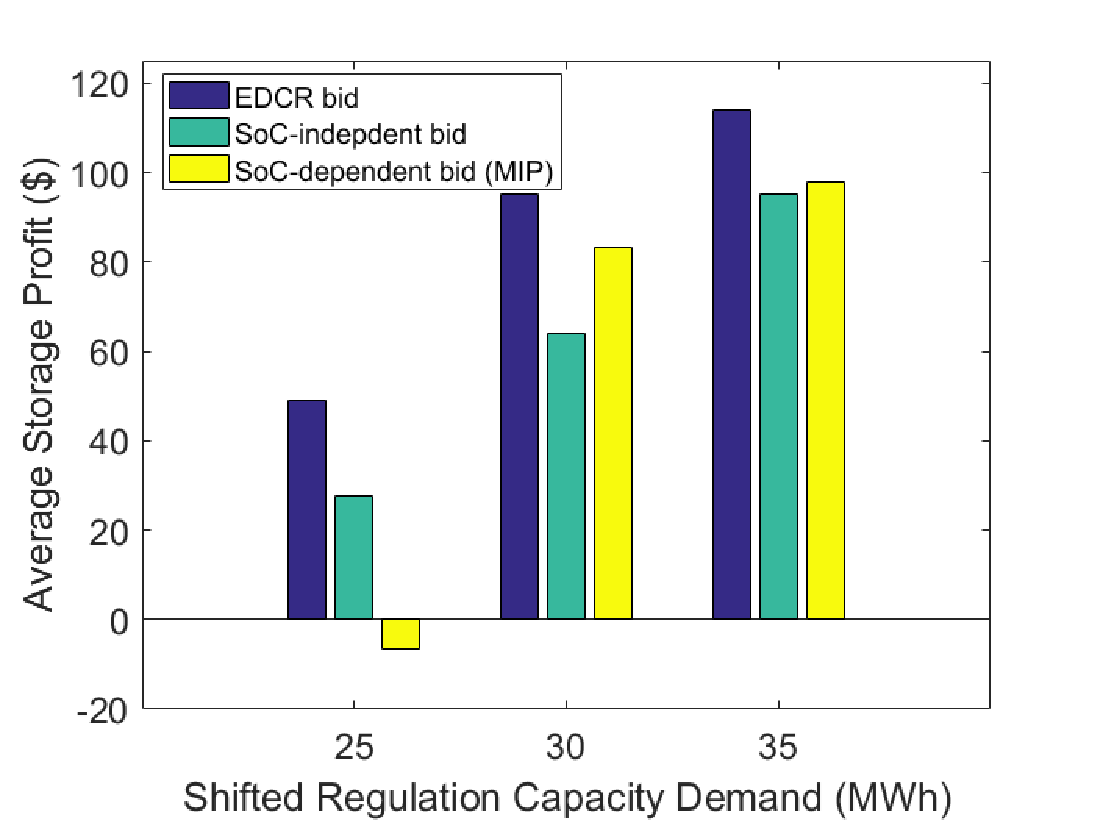}
   \includegraphics[scale=0.23]{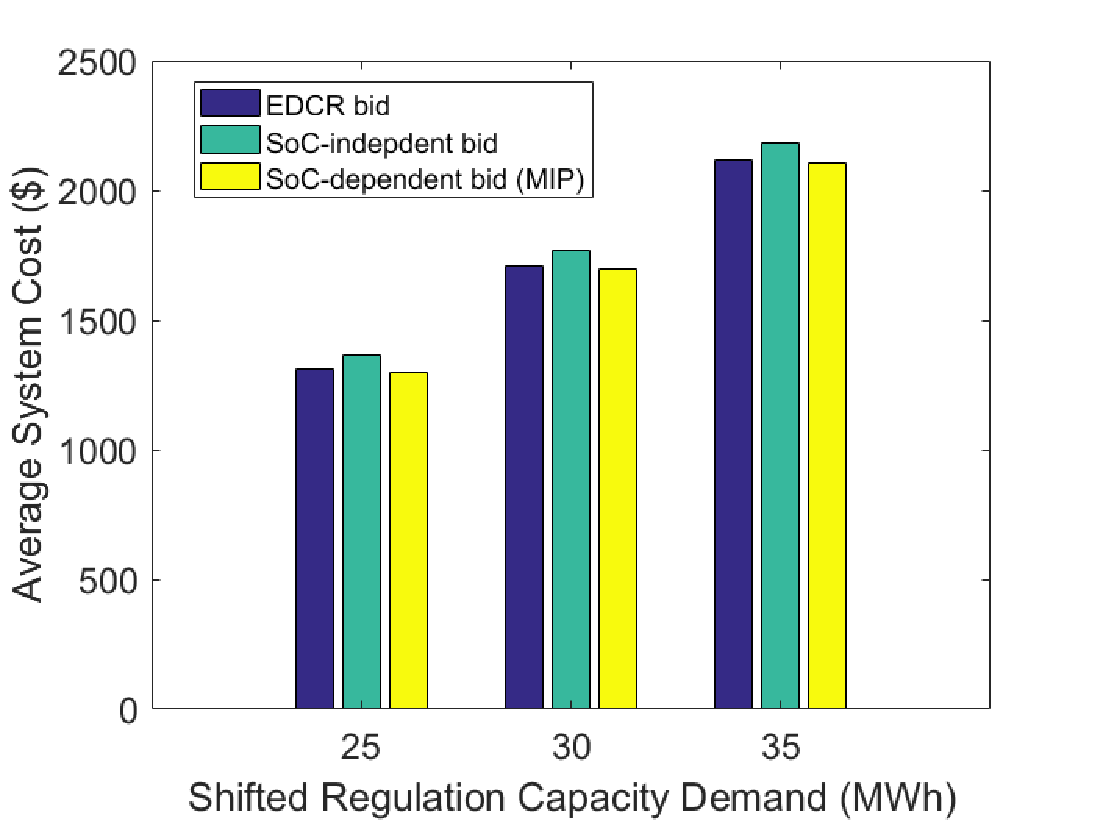}   \includegraphics[scale=0.23]{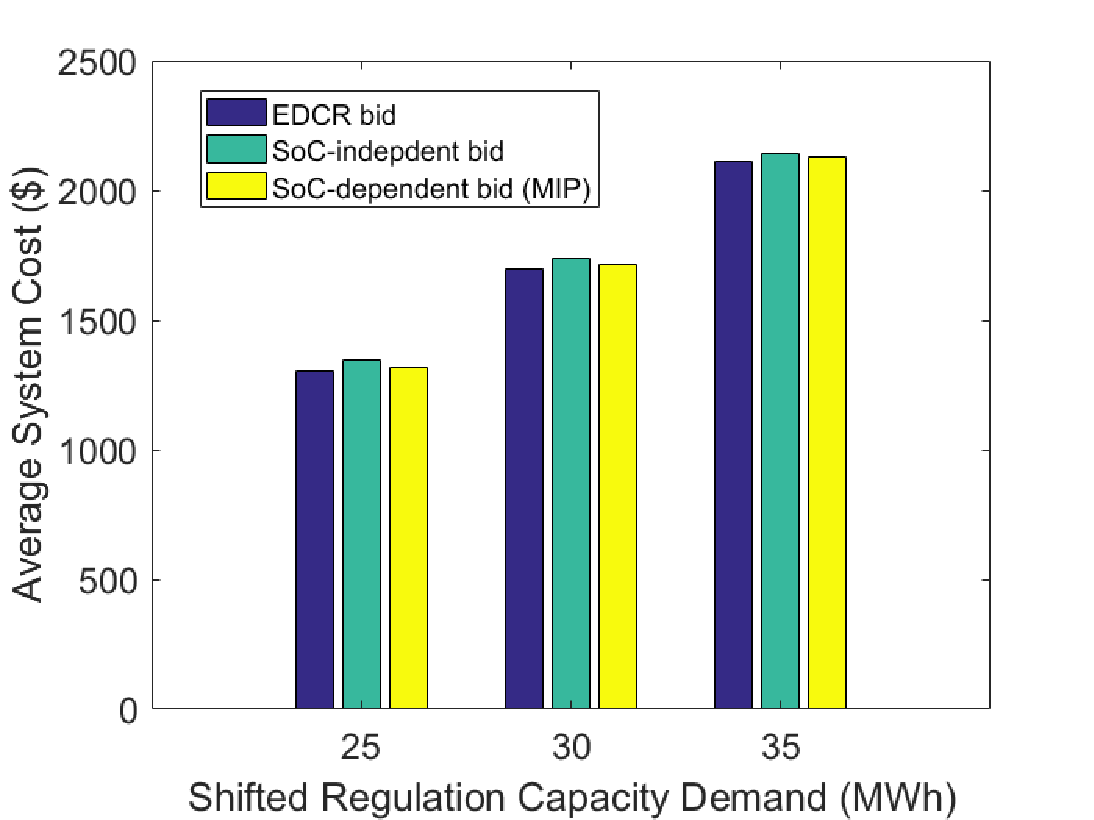} 
\caption{\scriptsize Simulation results. (Top left: storage profits in one-shot dispatch; top right: storage profits in rolling-window dispatch; bottom left: system cost in one-shot dispatch; bottom right: system cost in rolling-window dispatch).}
\label{fig:Oneshot Syscost}
\end{figure}

We conducted simulations for both one-shot and rolling-window dispatch. The former solved 24-T together; the latter solved 4-T in each rolling window and only implemented the first time interval. Shift demand quantities are added to the regulation capacity requirements $\xi^u_t$ and $\xi^d_t$, detailed in the appendix, to increase the demand from 25 MWh to 30 MWh and 35 MWh, respectively. This allows us to test cases with varying regulation demands. We computed the average evaluation metrics via averaging over 100 random scenarios and plotted them versus different levels of the shifted regulation demands, as shown in Fig.~\ref{fig:Oneshot Syscost}. Here are two conclusions.

First, storage accrued higher profits\footnote{Here we computed the true storage profit with \eqref{eq:true cost}.} with EDCR bids than that with SoC-independent bid. Right panel of Fig.~\ref{fig:ESbid} indicates that the increased profit resulted from the storage's more frequent charging and discharging behavior, which was enabled by its flexible SoC-dependent bid. As is shown by the top of Fig.~\ref{fig:Oneshot Syscost}, the storage profit under EDCR bid is 12.32-77.38\% higher than that under SoC-independent bid. In rolling-window dispatch, when the MIP simulated market clearing with storage bidding at the true SoC-dependent marginal cost, storage received less or even negative profits because of the nonconvexity pricing issue of MIP.

Second, when storage utilized EDCR bids, the system cost decreases by 1.38-4.17\% in comparison to the SoC-independent bid case (shown in the bottom of Fig.~\ref{fig:Oneshot Syscost}). 

\section{Conclusions}\label{sec:Conclusion}

The SoC-dependent bid provides BESS with greater flexibility to bid various parameters in the regulation market. To clear the SoC-dependent bid, we first introduce an intractable bilevel optimization for the regulation market.  It captures the worst SoC trajectories for the storage cost with nonconvex  subproblems. Sequentially, we propose the equal decremental-cost ratio (EDCR) condition to address nonconvexity. That way, we reformulate the bilevel optimization into a convex optimization, which can be computed efficiently in the real-time regulation market. 

In this paper, the storage only participates in the regulation market. Our future work \cite{ChenLiTong23SoC} plans to clear SoC-dependent bid in the energy-regulation joint market. Meanwhile, we anticipate to (i) generate SoC-dependent bids that ensure higher profitability compared to SoC-independent bids; (ii) conduct more practical simulations on larger-scale systems to better verify the effectiveness of our proposed method. 


{
\bibliographystyle{IEEEtran}
\bibliography{BIB}
}



\section{Appendix}
\label{sec:Appendix}

\subsection{Proof of Theorem~\ref{thm:ROCVX}}\label{sec:THM1}

We first introduce Lemma~\ref{lemma:Eq} and Lemma~\ref{lemma:ROsol}, and then we prove Theorem~\ref{thm:ROCVX}.

In Lemma~\ref{lemma:Eq}, we only analyze the storage cost for a single 15-min interval with index $t$. The initial SoC $e_t$ at the beginning of this 15-min interval is given. The SoC at the end of this 15-min interval is $e_{t+1}$. And we denote index $n$ and $m$ respectively below with $e_{t} \in {\cal E}_m$  and $e_{t+1} \in {\cal E}_n$.
\begin{lemma}[]\label{lemma:Eq}
If storage's bid-in parameters satisfy the equal decremental-cost ratio (EDCR) condition, then under Assumption~\ref{assume:single}, the  storage cost in (\ref{eq:SoCcostMile}) is piecewise linear convex given by
\begin{equation}
\begin{array}{lrl}\label{eq:EScostEDCR}
f^b(\mbf^u_t,\mbf^d_t,e_t)=\underset{j\in [K]}{\rm max}\{\alpha_j(e_t) + a^d_{j}\delta\mathbf{1}^\intercal \mbf^d_t+a^u_{j}\delta\mathbf{1}^\intercal \mbf^u_t\}\\
~~~~~~~~= a^d_{n}\delta\mathbf{1}^\intercal \mbf^d_t+a^u_{n}\delta\mathbf{1}^\intercal \mbf^u_t\\
~~~~~~~~~~~~~~+\begin{cases}
\sum_{k=m}^{n-1}\frac{\Delta a^d_{k}}{\eta}(E_{k+1}-e_t), & n>m \\
0, & m=n  \\
\sum_{k=n+1}^{m} \Delta a^u_{k-1} (E_{k}-e_t), & n<m
\end{cases}
\end{array}
\eeq
with $\alpha_j(e_t):=\sum_{k=1}^{j-1}\frac{\Delta a^d_k(E_{k+1}-E_1)}{\eta} + \frac{a^d_{j}(e_t-E_1)}{\eta}+h(e_t)$ and $h(e_t):=\sum_{i=1}^K\mathbbm{1}_{\{e_t\in \Ec_i\}}(\frac{a^d_i(E_1 - e_t)}{\eta} - \sum_{k=1}^{i-1}\frac{\Delta a^d_k (E_{k+1}-E_1)}{\eta})$.
\end{lemma} 

Proof: The proof follows the proof of Theorem 1 in \cite{ChenTong22arXivSoC}, with the distinction that we replace the SoC-dependent cost/benefit parameters from the energy market with the regulation capacity bid. 

Setting $\hat{e}_{\mbox{\tiny j+1}}=e_t+(\sum_{i=1}^{j}m^d_{i,t}\eta-\sum_{i=1}^{j}m^u_{i,t})\delta$, replacing $\frac{\Delta c^{\mbox{\tiny C}}_k}{\eta^c}$ and $\eta^{\mbox{\tiny D}}\Delta c^{\mbox{\tiny D}}_k$ in \cite{ChenTong22arXivSoC} with $-\frac{\Delta a^d_k}{\eta}$ and $\Delta a^u_k$ respectively. Noting that $\Delta a^d_k=-\Delta a^u_k \eta$, the detailed proof corresponds to the methodology outlined in our previous paper \cite{ChenTong22arXivSoC}. \QED

In our previous paper \cite{ChenTong22arXivSoC}, the EDCR condition for the energy market is proposed. If bid-in costs are derived from the value function of the stochastic storage optimization, as presented in \cite{ZhengXu22impact, ZhengXu22energy}, the resulting bids satisfy the EDCR condition. Here, in parallel, we establish Lemma~\ref{lemma:Eq} for the regulation market. \hfill 

With Lemma~\ref{lemma:Eq}, we achieve the analytical optimal solution for the robust storage cost \eqref{eq:ROmileage} below.
\begin{lemma}[]\label{lemma:ROsol}
If  storage's bid-in parameters satisfy the equal decremental-cost ratio (EDCR) condition, the optimal solution of \eqref{eq:ROmileage} has
\begin{align}
\label{eq:bindingCons}
    &\delta\mathbf{1}^\intercal \mbf^{u*} =  r^u_t,~\delta\mathbf{1}^\intercal \mbf^{d*} = r^d_t, \\ \label{lemma2}
    & f^*(r^u_t, r^d_t,e_t) = \underset{j\in [K]}{\rm max}\{\alpha_j(e_t) + a^d_{j} r^d_t+a^u_{j} r^u_t\}. \\\label{eq:objL}
    &~~~~~~~~= a^d_{n} r^d_t+a^u_{n} r^u_t\nn\\
&~~~~~~~~~~~~~~+\begin{cases}
\sum_{k=m}^{n-1}\frac{\Delta a^d_{k}}{\eta}(E_{k+1}-e_t), & n>m \\
0, & m=n  \\
\sum_{k=n+1}^{m} \Delta a^u_{k-1} (E_{k}-e_t). & n<m
\end{cases}
\end{align}

\end{lemma} 

Proof: First, we prove that to maximize the storage cost given the cleared quantities of regulation up capacity $r_t^u$ and regulation down capacity $r_t^d$, we have \eqref{eq:bindingCons}. Under Assumption~\ref{assume:single}, we know that $a^d_n \geq 0, a^u_n \geq 0, \delta =$ 4 s. Therefore, the objective function, with an equivalent form \eqref{eq:EScostEDCR} from  Lemma~\ref{lemma:Eq}, increases when $\mathbf{1}^\intercal \mbf^{u*}$  or $\mathbf{1}^\intercal \mbf^{d*}$ increases. We achieve the maximum objective value when \eqref{eq:bindingCons} is satisfied. Note that \eqref{eq:bindingCons} is always feasible for \eqref{eq:ROmileage}. One obvious feasible solution is $m^{u*}_{1,t}\delta=r^u_t,m^{d*}_{2,t}\delta=r^d_t$. Plugging \eqref{eq:bindingCons} into (\ref{eq:EScostEDCR}) results in the optimal objective value equal to \eqref{lemma2}. \QED

To solve the upper level optimization, we only need the optimal value rather than the optimal solution of the lower-level problem. And this is provided by Lemma~\ref{lemma:ROsol}. 
 \hfill 
 
Proof of Theorem~\ref{thm:ROCVX}:
We know that the initial SoC at $t=1$ is given, \ie $e_1=s$. Denote index $n$ and $m$ below respectively with initial SoC $e_{1} \in {\cal E}_m$  and the end-state SoC $e_{T+1} \in {\cal E}_n$. The SoC-dependent cost  \eqref{eq:robustcost} can be computed by 
\beq
\begin{aligned}\label{eq:proofTHM1}
F^*(&\pmb{r}^u,\pmb{r}^d; s):=\sum_{t=1}^{T}f^*(r^u_{t},r^d_{t},e_{t})\\
&\overset{(a)}{=} \sum_{t=1}^T f^b(\mbf^{u*}_t, \mbf^{d*}_t,e_t)\\
&\overset{(b)}{=}\sum_{t=1}^T \underset{j\in [K]}{\rm max}\{\alpha_j(e^*_t) + a^d_{j}\delta\mathbf{1}^\intercal \mbf^{d*}_t+a^u_{j}\delta\mathbf{1}^\intercal \mbf^{u*}_t\}\\
&\overset{(c)}{=}  \sum_{t=1}^T \underset{j\in [K]}{\rm max}\{\alpha_j(e_t) +  a^d_{j}r_t^d+a^u_{j} r_t^u\}\\
    &\overset{(d)}{=} \sum_{t=1}^T a^d_{n}r^d_t+a^u_{n}r^u_t\\
&~~~~~~~~+\begin{cases}
\sum_{k=m}^{n-1}\frac{\Delta a^d_{k}}{\eta}(E_{k+1}-s), & n>m \\
0, & m=n  \\
\sum_{k=n+1}^{m} \Delta a^u_{k-1} (E_{k}-s), & n<m
\end{cases} \\
&\overset{(e)}{=}\underset{j\in [K]}{\rm max}\{\alpha_j(s) + \sum_{t=1}^T(a^d_{j}r_t^d+a^u_{j}r_t^u)\},
\end{aligned}
\eeq
where (a) comes from the definition in \eqref{eq:ROmileage}, (b) comes from \eqref{eq:EScostEDCR} in Lemma~\ref{lemma:Eq}, (c) comes from  \eqref{eq:bindingCons} and \eqref{lemma2} in Lemma~\ref{lemma:ROsol}.  (d) relies on \eqref{eq:objL} and can be proved by the same induction method in the proof of Theorem 1 in \cite{ChenTong22arXivSoC}. Note that, a difference between the proof of \cite{ChenTong22arXivSoC} and this paper is to use different SoC evolving constraint $
e_{t+1}=e_{t} + r^d_{t}\eta-r^u_{t}$ from \eqref{eq:SoCevolveReg}, which allows two-side regulation capacity cleared. Although the proof in \cite{ChenTong22arXivSoC} is written based on one-side clearing of energy market, the same induction works here for two-side clearing of regulation capacity. And (e) follows the proof of Theorem 1 in \cite{ChenTong22arXivSoC} showing that 
\beq
n=\underset{j\in [K]}{\rm arg~max}\{\alpha_j(s) + \sum_{t=1}^T(a^d_{j}r_t^d+a^u_{j}r_t^u)\}.
\eeq

The last step is to prove that \eqref{eq:ESROcost} is a convex function by showing it uses operations preserving convexity.
We know  $\alpha_j(s)$ is a constant, \eqref{eq:ESROcost} is adopting pointwise maximum operation over linear functions, so \eqref{eq:ESROcost} is  convex. \QED
\subsection{Proof of Proposition~\ref{Prop:Simultaneous}}\label{sec:prop1}
We complete the dual variables for \eqref{eq:NONCVX_REG_UP} below with the EDCR condition satisfied\footnote{we substitute the storage cost in model (\ref{eq:obj}) with \eqref{eq:ESROcost}}, and then write the proof.

\begin{equation} \label{eq:NONCVX_REGAll}
\begin{array}{lrl}
& \underset{\substack{\{\pmb{r}_{i}^u,\pmb{r}_{i}^d,\pmb{e}_{i},\\ \gbf_i^e, \gbf_i^u, \gbf_i^d \in \Rbf^+ \}}}{\rm minimize} & \sum_{i=1}^{M} (h_i(\gbf^e_i, \gbf_i^{u}, \gbf_i^{d})+ \tilde{F}_{i}(\rbf^u_{i}, \rbf^d_{i};s_i))\\
& \mbox{subject to}& \forall t\in [T], \forall i\in [M],\\
&& \pmb{S} (\pmb{g}^e[t]-\pmb{d}[t]) \le \pmb{q}\\
&& {\bf 1}^\intercal(\pmb{g}^e[t]-\pmb{d}[t]) = 0\\
&\phi_{it}: & e_{it} + r^d_{it}\eta_i-r^u_{it}=e_{i(t+1)},\\
&\beta^u_{t}: & \sum_i^M (g_{it}^{u} + r_{it}^u) \geq \xi_t^u, \\
&\beta^d_{t}: & \sum_i^M (g_{it}^{d} + r_{it}^d) \geq \xi_t^d, \\
&& e_{i1}=s_i,\\
&& g^e_{it}+g^u_{it}\leq \bar{g}_i,\\
&& \underline{g}_i \le g^e_{it}-g^d_{it}, \\
&(\underline{\rho}^d_{it},\bar{\rho}^d_{it}): & 0\le r^d_{it}\le \bar{r}^d_i,\\
&(\underline{\rho}^u_{it},\bar{\rho}^u_{it}): & 0\le r^u_{it}\le \bar{r}^u_i,\\
&\omega^d_{it}: & e_{it}+ r^d_{it}\eta_i \leq \bar{E}_i,\\
&\omega^u_{it}: & \underline{E}_i \le e_{it}- r^u_{it}.\\
\end{array}
\end{equation}

Proof: Assume for the sake of contradiction that there exists an optimal solution in which storage $i$ simultaneously offers both regulation up and regulation down capacities at time interval $t$, \ie $r^u_{it}>0, r^d_{it}>0$. After the EDCR condition is satisfied, the regulation market clearing model (\ref{eq:NONCVX_REG_UP}) exhibits convexity with respect to the subdifferentiable objective function \eqref{eq:ESROcost}. With the KKT conditions, there exist subgradients $\kappa^u_i\in \frac{\partial}{\partial r^u_{it}}  \tilde{F}_i(\rbf^u_i, \rbf^d_i; s_i)$ and $\kappa^d_i \in  \frac{\partial}{\partial r^d_{it}}  \tilde{F}_i(\rbf^u_i, \rbf^d_i; s_i)$ such that
\beq\label{eq:EDCRKKT}
\begin{cases} 
 \kappa^u_i+(-\phi^*_{it}+\omega^{u*}_{it})-\beta^{u*}_t-\underline{\rho}_{it}^{u*}+\bar{\rho}_{it}^{u*}=0, & \\
 \kappa^d_i+(\phi^*_{it}+\omega^{d*}_{it})\eta_i-\beta^{d*}_t-\underline{\rho}_{it}^{d*}+\bar{\rho}_{it}^{d*} =0,& 
\end{cases}
\eeq
\beq
\begin{aligned}\label{eq:contradict}
\Rightarrow &\kappa^u_{i}-\beta^{u*}_t-\underline{\rho}_{it}^{u*}+\bar{\rho}_{it}^{u*}+\frac{\kappa^d_i-\beta^{d*}_t-\underline{\rho}_{it}^{d*}+\bar{\rho}_{it}^{d*}}{\eta_i}\\
&~~~~~~~~~~~~~~~~~~~~~~~~~~~~~~+\omega^{u*}_{it}+\omega^{d*}_{it}=0,
\end{aligned}
\eeq
where $\bar{\rho}_{it}^{u*}\geq 0, \bar{\rho}_{it}^{d*}\geq 0, \omega^{u*}_{it}\geq 0, \omega^{d*}_{it}\geq 0$, and  we have $\underline{\rho}_{it}^{u*}=0, \underline{\rho}_{it}^{d*}=0$ from the complementary slackness conditions. The subgradients $\kappa^u_i$ and $\kappa^d_i$ are equal to $a^u_{ik_1}$ and $a^d_{ik_2}$ respectively for some $k_1,k_2\in [K]$. 
In Proposition~\ref{Prop:Simultaneous}, restrictions are only imposed on $a_K^u$ and $a_1^d$ due to the monotonicity of the SoC-dependent bid under Assumption~\ref{assume:single}. So we have,  $\forall k_1,k_2\in[K]$,
\beq
a_{iK}^u\eta_i+a_{i1}^d>\beta^{u*}_t\eta_i+\beta^{d*}_t \Rightarrow a_{ik_1}^u\eta_i+a_{ik_2}^d>\beta^{u*}_t\eta_i+\beta^{d*}_t.\nn\eeq

Therefore, we have 
$\kappa^u_i \eta_i+\kappa^d_i>\beta^{u*}_t\eta_i+\beta^{d*}_t \Rightarrow $
\beq
\kappa^u_{i}-\beta^{u*}_t+\bar{\rho}_{it}^{u*}+\frac{\kappa^d_i-\beta^{d*}_t+\bar{\rho}_{it}^{d*}}{\eta_i}+\omega^{u*}_{it}+\omega^{d*}_{it}>0,
\eeq
leading to a contradiction for \eqref{eq:contradict}. \QED

\subsection{Simulation parameters}

\begin{figure}[!htb]
   \centering
   \vspace{-0.1in}
   \includegraphics[scale=0.5]{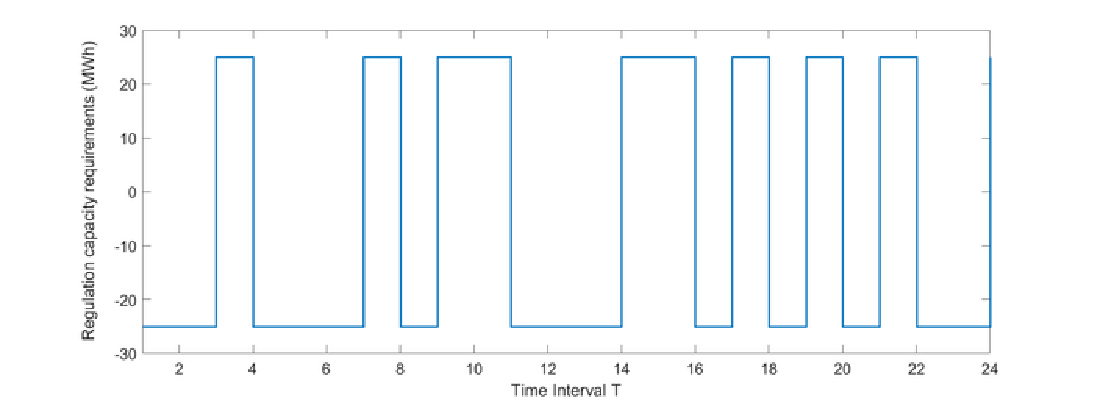} 
\caption{\scriptsize Regulation capacity demand in the simulation. Positive values on the y-axis represent $\xi^u_t$, while negative values equal $-\xi^d_t$ for different intervals.}
\label{fig:RegRequire}
\end{figure}

According to the real-time regulation market results of PJM \cite{PJM:DataMiner}, the required amount of  regulation capacities remains stable, while the clear price for regulation capacities exhibits volatility. To simulate such a phenomenon in our case studies, we set the regulation capacity demand to be constant over time, as shown in Fig.~\ref{fig:RegRequire}. In the figure, positive values represent regulation up requirements, while negative values represent regulation down requirements. Meanwhile, we introduce fluctuations in the output of the wind generator to simulate the volatility of the clearing price for regulation capacity.

\begin{figure}[!htb]
   \centering
   \vspace{-0.1in}
   \includegraphics[scale=0.28]{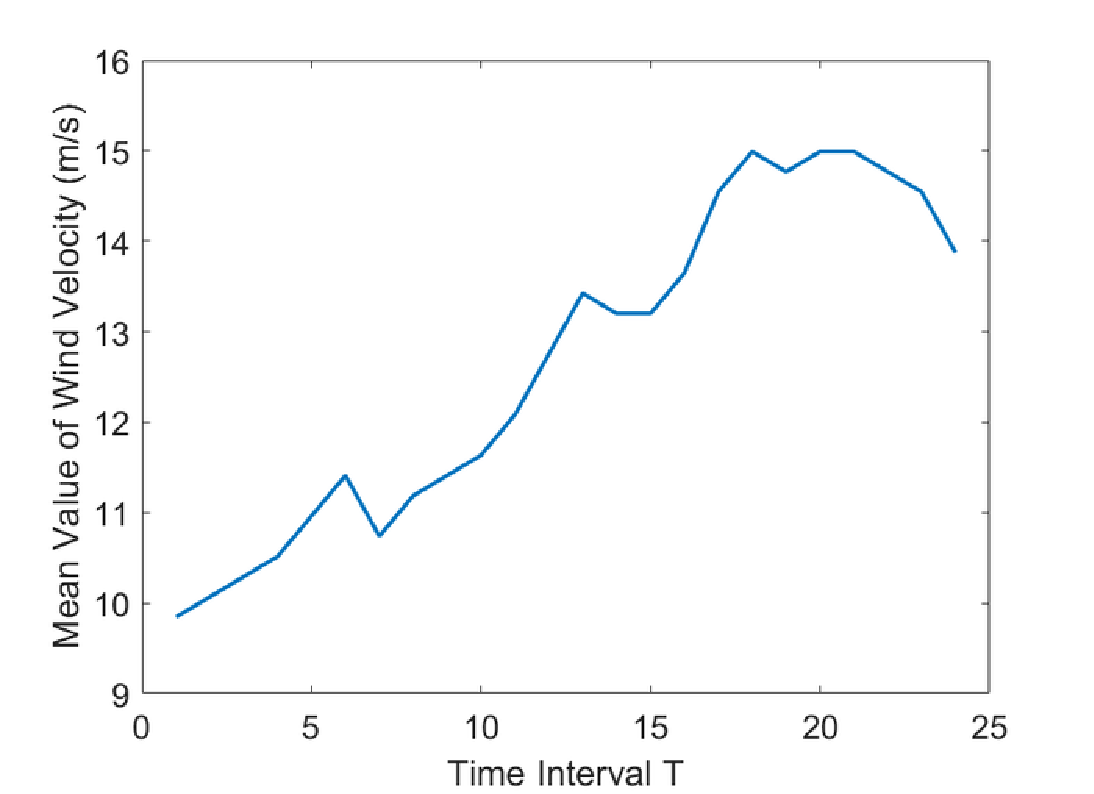}
\caption{\scriptsize Mean value of wind velocity.}
\label{fig:WindSolar}
\end{figure}

In the simulation, we generate wind velocity with Gaussian distribution volatility. Given the mean value of wind velocity $\pmb{\mu}$ for each interval during the 24-T period, as shown in Fig.~\ref{fig:WindSolar}, the wind velocity at time $t$ follows ${\cal N}(\mu_t, 5)$.


\subsection{Mixed integer SoC-dependent regulation market model}\label{sec:MIPheuristic}

The challenge in solving \eqref{eq:NONCVX_REG_UP} lies in clearing regulation capacities when regulation mileages following AGC are unknown. Here, to avoid addressing a complicated bilevel model, we provide a heuristic approach to approximate the worst-case SoC-dependent storage cost in \eqref{eq:ROmileage}. Such a heuristic is accurate when unidirectional regulation capacities are cleared.
\beq
\begin{aligned}\label{eq:ES_Scost_Approx}
\check{f}(r^u_{it},r^d_{it},e_{it}):=\min\{&f^u(\frac{r^u_{it}}{\delta},e_{it})+f^d(\frac{r^d_{it}}{\delta}, e_{it} - \frac{r^u_{it}}{\delta}), \\&f^d(\frac{r^d_{it}}{\delta}, e_{it}) +f^u(\frac{r^u_{it}}{\delta},e_{it} + \frac{r^d_{it}}{\delta}) \},
\end{aligned}
\eeq 
where $f^u$ and  $f^d$ are the SoC-dependent cost for regulation up and regulation down respectively defined under \eqref{eq:SoCcostMile} in the main text.

The intuition of such approximation is to consider the worst type of SoC trajectories within the 15-minute interval of time $t$. This involves evaluating two possibilities: first, fully utilizing the cleared regulation up capacity followed by the cleared regulation down capacity, and second, the reverse order - starting with the cleared regulation down capacity and then using up the cleared regulation up capacity. \eqref{eq:ES_Scost_Approx} accurately capture the worst-case SoC-dependent cost for storage when only one-side regulation market is cleared at each time interval.

By approximating $F^*_{i}(\rbf^u_{i}, \rbf^d_{i};s_i)$ in \eqref{eq:NONCVX_REG_UP} with $\sum_{t=1}^{T}\check{f}(r^u_{it},r^d_{it},e_{it})$, we propose a MIP to solve the energy-regulation co-optimization.

We first introduce the following notations specifically used in this MIP:
\beq
\begin{array}{l}
\pmb{r}^u_{ik} = (r^u_{i,k,1},...,r^u_{i,k,t}, ..., r^u_{i,k,T} ),\\
\pmb{r}^u_{it} = (r^u_{i,1,t},...,r^u_{i,k,t}, ..., r^u_{i,K,t} ),\\
\pmb{e}_{it} = (e_{i,1,t},...,e_{i,k,t}, ..., e_{i,K,t} ),\nn
\end{array}
\eeq
where $r^u_{i,k,t}$ represents the regulation up capacity of storage $i$ in time interval $t$ from SoC segment $k$, and $e_{i,k,t}$ denotes the state of energy stored in segment $k$ at the beginning of interval $t$.

In this optimization problem, the domain for  variables is $\hat{\Omega} = \{\pmb{r}_{ik}^u,\pmb{r}_{ik}^d, \gbf_i^e, \gbf_i^u, \gbf_i^d \in \Rbf_+^T, I'_t, I''_t \in \Bbf,\ubf_{it},  \ubf'_{it},  \ubf''_{it}\in \Bbf^{K}\}$, and the $T$-interval SoC-dependent storage cost at the segment $k$ is given by 
\beq
\varphi_{ik}(\rbf^u_{ik}, \rbf^d_{ik}) =  a_{ik}^u{\bf 1}^\intercal\pmb{r}^u_{ik} + a_{ik}^d{\bf 1}^\intercal\pmb{r}_{ik}^d.\eeq

The formulation of the model is as follows.
\begin{subequations} \label{eq:MILP_REG}
    \begin{align}
& \underset{ \hat{\Omega} }{\rm minimize} && \sum_{i=1}^{M} (h_i(\gbf^e_i, \gbf_i^{u}, \gbf_i^{d})+\sum_{k=1}^{K} \varphi_{ik}(\rbf^u_{ik}, \rbf^d_{ik})) \\
& \mbox{subject to}&& \forall t\in [T], \forall i\in [M],\nn\\
&&& \pmb{S} (\pmb{g}^e[t]  -\pmb{d}[t]) \le \pmb{q}, \\
&&& {\bf 1}^\intercal(\pmb{g}^e[t] -\pmb{d}[t]) = 0,\\
&&& \sum_i^M (g_{it}^{u} + {\bf 1}^\intercal\rbf_{it}^u) \geq \xi_t^u, \\
&&& \sum_i^M (g_{it}^{d} + {\bf 1}^\intercal\rbf_{it}^d) \geq \xi_t^d, \\
& && \ebf_{it} +  \rbf^d_{it}\eta_i-  \rbf^u_{it}=\ebf_{i(t+1)},\\
&&& \ebf_{it} +  \rbf^d_{it}\eta_i \leq \bar{E}_i, ~~\underline{E}_i \le \ebf_{it}  -  \rbf^u_{it},\\
&&& \ebf_{i1}=\sbf_i, ~~\underline{\Ebf}_i\le e_{it} \le \bar{\Ebf}_i,\\
&&& g^e_{it}+g^u_{it}\leq \bar{g}_i, ~~\underline{g}_i \le g^e_{it}-g^d_{it}, \\
&&& \ebf_{it}+ \rbf^{d'}_{it}\eta_i = \ebf'_{it}, ~~\ebf'_{it}- \rbf^{u'}_{it} = \ebf_{i(t+1)},\\
&&& \ebf_{it}-  \rbf^{u''}_{it} = \ebf''_{it}, ~~\ebf''_{it}+ \rbf^{d''}_{it}\eta_i = \ebf_{i(t+1)},\\
&&& J(I'_t, \rbf^{u'}_{i,t},\rbf^{u}_{it}, \rbf^{d'}_{it}, \rbf^{d}_{it}) \le 0,\\
&&& J(I''_t, \rbf^{u''}_{it},\rbf^{u}_{it}, ~\rbf^{d''}_{it}, \rbf^{d}_{it}) \le 0,\\
&&& I'_t + I''_t =1,\\
&&& \phi(\ebf_{it}, \ubf_{it}) \le 0,  \phi(\ebf'_{it}, \ubf'_{it}) \le 0,   \phi(\ebf''_{it}, \ubf''_{it}) \le 0.\nn
    \end{align}
\end{subequations}

In addition to the constraints introduced in model (\ref{eq:NONCVX_REG_UP}), route selection functions, denoted as $ J(I'_t, \rbf^{u'}_{i,t},\rbf^{u}_{i,t}, \rbf^{d'}_{i,t}, \rbf^{d}_{i,t}) \le 0 $, are used to select between the two possible SoC trajectories considered in (\ref{eq:ES_Scost_Approx}), with details as follows.
\beq
\begin{array}{l}
J(\cdot) = \begin{pmatrix} -M I'_t {\bf 1}+ \rbf^{d'}_{i,t} - \rbf^{d}_{i,t} \\ -M I'_t {\bf 1}+ \rbf^{u'}_{i,t} - \rbf^{u}_{i,t} \end{pmatrix} \le  0,
\end{array}
\eeq
where $I'_t$ is a binary variable, and $M$ is an assigned large number. By writing a vector no larger than zero, we mean each element of the vector is no less than zero.

Ignoring the time index $t$ and unit index $i$, we also provide the details for the last constraint $ \phi((e_{k}), (u_{k})) \le 0$. These constraints ensure the SoC segment logic for the piecewise model \cite{ZhengXu22energy}, $\forall k \in \{2,...,K-1\}$
\beq
\begin{array}{l}
\phi(\cdot) = \begin{pmatrix}
-M(1-u_{1}) - e_{1}\\  
e_{1} - E_1\\
- e_{K}\\
e_{K}- Mu_{K-1}\\
...\\
(E_k-E_{k-1})u_{k} - e_{k}\\
e_{k} - (E_k-E_{k-1})u_{k-1}\\
...\\
\end{pmatrix} \le  0.
\end{array}
\eeq




\subsection{Linear program reformulation under EDCR condition}\label{sec:LP}
The EDCR bid convexifies the bilevel non-convex problem, enabling its reformulation as the following standard linear program.
\begin{subequations}
\label{eq:SdLinear}
\begin{align}
& \underset{\substack{\{\gbf_i^e, \gbf_i^u, \gbf_i^d,\in \Rbf^T_+,\\\pmb{r}_{i}^u,\pmb{r}_{i}^d \in \Rbf^T_+,\pmb{\psi}\in \Rbf^M \}}}{\rm minimize} && \sum_{i=1}^{M} (h_i(\gbf^e_i, \gbf_i^{u}, \gbf_i^{d})+ \psi_{i}) \\
& \mbox{subject to}&& \forall t\in [T], \forall i\in [M],\forall j\in [K],\nn\\
&&& \eqref{eq:PFlimit}-\eqref{eq:SoCcons2},\nn \\
&&& \psi_{i}\geq \alpha_{ij}(s) + \sum_{t=1}^T(a^d_{ij}r_{it}^d+a^u_{ij}r_{it}^u),
    \end{align}
\end{subequations}
where $\psi_{i}$ denotes the worst-case SoC-dependent cost of storage $i$, which is the optimal value of lower level model \eqref{eq:ROmileage}. As introduced in Theorem~\ref{thm:ROCVX}, $\alpha_{ij}(s) + \sum_{t=1}^T(a^d_{ij}r_{it}^d+a^u_{ij}r_{it}^u)$ represents the cost of piece $j$ of storage $i$. For simplicity, we assume  generation cost $h_i$ here is a linear function. In practice, it can be convex piecewise linear function and we can reformulate it into linear program. 

\end{document}